\documentclass[12pt]{article}
\usepackage{amsmath,amssymb}
\newcommand{\eqr}[1]{(\ref{#1})}
\newcommand{\ds}{\displaystyle}
\newtheorem{theorem}{Theorem}%\usepackage{amsmath}
\numberwithin{equation}{section}

\newcommand{\ri}{\right}
\newcommand{\lf}{\left}    
\newcommand{\beq}{\begin{equation}}
\newcommand{\eeq}{\end{equation}}
\newcommand{\bea}{\begin{eqnarray}}
\newcommand{\eea}{\end{eqnarray}}
\newcommand{\de}{\partial}
\newcommand{\nn}{\nonumber}
\newcommand{\TMP}[1]{{\sl Theor.\ Math.\
Phys.}\ {\bf #1}}
\begin{document}
\today

\begin{center}
\bigskip

{\bf Point Symmetries of Generalized Toda Field Theories}

\bigskip S. Lafortune\footnote{E-mail:lafortus@CRM.UMontreal.CA}
  and P.   
Winternitz\footnote{E-mail:
wintern@CRM.UMontreal.CA}    

{\it Centre de Recherches Math\'{e}matiques, Universit\'{e} de   
Montr\'{e}al,
C.P. 6128, Succ. Centre-Ville, Montr\'{e}al, Qu\'ebec, H3C 3J7,   
Canada }

L. Martina  \footnote{E-mail:
martina@le.infn.it}    

{\it Dipartimento di Fisica dell'Universit\`a and INFN
Sezione di    
 Lecce, C.P. 193, 73100 Lecce, Italy}
\end{center}

\vspace{2in}

\medskip Abstract \qquad {\it A class of two-dimensional field theories with   
exponential interactions   
is introduced. The interaction depends on two ``coupling'' matrices and is   
sufficiently general to   
include all
Toda field theories existing in the literature. Lie point symmetries of these   
theories are found for an   
infinite,   
semi-infinite and finite number of fields. Special attention is
 accorded to conformal invariance and its   
breaking.}

\newpage    

\section{Introduction\protect\bigskip}

The purpose of this article is to investigate the Lie point   
symmetries    
of a
large class of  ``generalized Toda field theories''. The class is    
characterized
by the equation
\begin{equation}
u_{n,xy}= F_{n},\hspace{1cm} F_{n}=    \sum_{m=n-n_{1}}^{n+n_{2}}   
K_{nm} \exp    
\left( {{\sum_{l=    m-n_{3}}^{m+n_{4}} H_{ml} u_{l}}}\right)    
\label{1.1}
\end{equation}
where $K$ and $H$ are some real constant matrices and $n_{1},\dots   
,n_{4}$ are      
some
finite non-negative integers. The range of $n$ may be infinite, semi-infinite
 or   finite,      
hence
the matrices $K$ and $H$ may also be infinite, semi-infinite,
 or finite.   
     
If
the range of $n$ is finite, $K$ and $H$ may be rectangular, not      
necessarily
square. We assume that all the rows in $H$ are different, that
$H$    
contains
no zero rows and $K$
no zero columns. In all the cases we assume that the range of      
the
interaction on the right hand side of eq. \eqr{1.1} is finite,   
hence the   
finite
summation limits in both sums. ``Generalized Toda lattices''   
are obtained from eq. \eqr{1.1} by   
symmetry    
reduction,
using translational invariance, i.e. restricting to solutions of the   
form    
$u_n(x,y) =     
w_n(t)$ where $t = x+\lambda y$.

Toda lattices and their generalisations, Toda field theories,   
represent   
one of    
the most
interesting, rich and fruitful developments in the realm of
 completely    
integrable
systems.
The original Toda lattice was introduced by M. Toda \cite{1,2}
who found   
analytical
solitons and periodic solutions in a discrete lattice with an   
exponential    
potential
involving nearest neighbour interactions. It was also found that
 the   
Toda    
lattice
admits a Lax representation and all the usual attributes of   
integrability    
\cite{3,4}.
The Toda lattice was generalized to integrable lattices   
related   
to the    
root systems    
of simple Lie algebras \cite{5} - \cite{8}. The considered   
lattices   
can be    
finite, infinite,
semi-infinite, or periodic.

The attractive features of Toda lattices have been generalized    
to two space dimensions in several different ways   
\cite{9} - \cite{19}.

All of them can be recovered from eq. \eqr{1.1} by specifying the   
matrices $K$    
and $H$.
Thus, the Mikhailov-Fordy-Gibbons field theories \cite{9,10}   
(for   
infinitely    
many fields)
\begin{equation}
u_{n,xy} = e^{u_{n-1}-u_n}-e^{u_n-u_{n+1}}
\label{1.2}
\end{equation}
are obtained by putting $H_{nn-1}  -H_{nn}=  1$,   
$K_{nn}=  -K_{nn+1}=  1$ and all other    
components to zero. A class of Toda field theories    
\begin{equation}
u_{n,xy}=  \sum_{m=  n-n_1}^{n+n_2}K_{nm} e^{u_m},
\label{1.3}
\end{equation}
studied by Leznov and Saveliev \cite{12,13}, Olive, Turok and   
others   
\cite{14} - \cite{17}    
(usually for a finite number of fields $u_n$) are obtained by   
setting   
$H=  I$ and    
taking $K$    
to be the Cartan matrix of a semisimple Lie algebra (or an
affine one).

A further class of Toda field theories, also studied by Leznov and   
Saveliev    
\cite{13,14}, by  Bilal and Gervais \cite{17}, and Babelon and   
Bonora   
\cite{18} (for a    
finite number of fields)
can be written as
\begin{equation}
u_{n,xy}=  \exp{\sum_{l=  m-n_3}^{m+n_4}H_{nl}u_l}
\label{1.4}
\end{equation}
and is obtained by taking $K=  I$ and $H$ as a Cartan matrix.

In this article we will be interested in point symmetries of the   
system    
\eqr{1.1},    
rather than in questions of integrability, or explicit solutions.   
The   
symmetries    
we are interested in will include conformal invariance, whenever   
it is   
present,    
and gauge
invariance, not however higher, or generalized symmetries, be they   
local, or    
not.

In Section 2 we consider infinite Toda field theories, i.e.   
take    
$-\infty<n<\infty$.    
In this case eq. \eqr{1.1} can be viewed as a   
differential-difference   
equation.    
Continuous    
Lie symmetries of such equations have been studied using several   
different    
approaches \cite{20} - \cite{29}. We shall follow that of Ref.   
\cite{20} - \cite{24},   
using both    
the    
``intrinsic method'' and the ``differential equation method''
 \cite{21}.

In Section 3 we turn to finite Toda field theories, when we   
have $1\leq   
n\leq    
N<\infty$ in    
eq. \eqr{1.1}. Eq. \eqr{1.1} in this case represents a system   
of $N$    
differential equations and its
point symmetries can be obtained in a standard manner   
\cite{30,31}. We first    
obtain
general results, then specify the matrices $H$ and $K$ in several    
different ways.

Section 4 is devoted to semi-infinite
Toda field theories, i.e. $0\leq n<\infty$. Again we first obtain
general results, then specify the matrices $H$ and $K$,
inforcing the cut-off at $n=  0$ in several different ways.

Some conclusions are drawn in Section 5.

\section{{\ Symmetries of Generalized }$\infty -${Toda Field   
Theories }}

\subsection{\protect\bigskip \noindent {\sl General Results}      
\protect\medskip
\noindent}

Let us consider eq. \eqr{1.1} with $n$ in the range $-\infty   
<n<\infty   
$.      
We
follow the ``differential equation method'' described in   
Ref.\cite{21} and
look for transformations of the form     
\begin{equation}
{\tilde{\vec{x}}}=    \Lambda _{g}(\vec{x},\{u_{k}\}),
\;\;\tilde{u}_{n}   
=    
 \Omega    
_{g}(\vec{x}
,n,\{u_{k}\}),\;\;\tilde{n}=n,  \label{3}
\end{equation}
where we have used the notation $\vec{x}\equiv      
(x,y)$, ${\tilde{\vec{x}} \equiv  (\tilde{x},%
\tilde{y})}$,     
taking solutions of eq. \eqr{1.1} into solutions. The notation   
$\{u_{k}\}$
indicates that the new variables can depend on all the fields $%
\{u_{k}\}_{k\in {\bf Z}}$.

The Lie group transformation \eqr{3} is generated by a Lie algebra   
of      
vector
fields of the form     
\begin{equation}
\hat{v}=    \xi (x,y,\{u_{k}\})\partial _{x}+\eta   
(x,y,\{u_{k}\})\partial
_{y}+\sum_{j=    -\infty }^{\infty }\phi _{j}(x,y,\{u_{k}\})
\partial_{u_{j}}.
\label{4}
\end{equation}
The prolongation of this vector field is constructed in the same   
manner    
as
for differential equations \cite{30,31} (albeit an infinite 
system of them). For   
a general equation   
of the     
form     
\begin{equation}
E_{n}=    u_{n,xy}-F_{n}(x,y,\{u_{k}\})=    0,  \label{5}
\end{equation}
we require     
\begin{equation}    
pr^{\left( 2\right) }\hat{v}E_{n}\mid _{E_{n}=    0}=    0.    
\label{6}
\end{equation}
It was shown quite generally \cite{21} that for eq \eqr{5}   
with $F_{n}$    
any
sufficiently smooth function depending on at least one function $u_k$, 
$k\neq  n$,   
the vector field   
\eqr{4}   
satisfying      
eq. \eqr{6}
will have the form
\begin{equation}
\xi  =   \xi (x),\ \ \eta =    \eta (y),\ \ \phi _{n}=     
 \sum_{k= -\infty      
}^{\infty
}A_{nk}u_{k}+B_{n}(x,y) ,  \label{7}
\end{equation}
where $A=    \{A_{n\alpha }\}$ is a constant (infinite)      
matrix. The
functions in eq. \eqr{7} must satisfy a remaining determining   
equation,
namely     
\begin{equation}
\begin{array}{l}
\displaystyle{B_{n,x y}-(\xi_{x}+\eta _{y}) F_{n}+
{\sum_{\alpha =    -\infty}^{\infty} }A_{n\alpha }F_{\alpha }
-\xi F_{n,x}-\eta F_{n,y}} \\  \\
\displaystyle{\hspace{0.32in}-
{\sum_{ \alpha =    -\infty}^{\infty} }
\left(    
{\sum_{\beta =    -\infty}^{ \infty}} A_{\alpha
\beta }u_{\beta }+B_{\alpha }\right) F_{n,u_{\alpha }}=    0},
\end{array}
\label{det-eq}\end{equation}
where $F_{n,u_{\alpha }}$ is the derivative of $F_{n}$ with
respect to the variable $u_{\alpha }$.

Let us now specify the function $F_{n}$ to be a sum of   
exponentials as     
in eq.   
\eqr{1.1}. There are three types of terms in eq. \eqr{det-eq}:     
those independent      
of $%
u_{n}$, linear in $u_{n}$ times exponentials and pure   
exponentials. Each
type of term must vanish separately. Since $H$ has no zero rows
 we get the determining   
equations   
\begin{equation}
B_{n,xy}=    0,  \label{9}
\end{equation}
\begin{equation}
{\sum_{\alpha =    -\infty}^{\infty } } A_{\alpha
m}F_{n,u_{\alpha }}=    0,  \label{10}
\end{equation}
\begin{equation}
-(\xi _{x}+\eta _{y})F_{n}+
{\sum_{\alpha=    -\infty }^{\infty } } A_{n\alpha }F_{\alpha }-
{\sum_{\alpha=    -\infty }^{\infty } } B_{\alpha }
F_{n,u_{\alpha }}=
0.       
\label{11}
\end{equation}
Eq. \eqr{10} can be rewritten as     
\begin{equation}
\sum_{\alpha \beta }K_{n\beta }H_{\beta \alpha }
A_{\alpha m}\exp \left(
\sum_{\gamma }H_{\beta \gamma }u_{\gamma }\right) =      
0.  \label{12}
\end{equation}
All exponentials in eq. \eqr{12} are linearly independent (since all   
rows      
in $H$ are different), so the equation must hold for each $\beta $      
separately
and the exponentials can be dropped. Moreover, the factor   
$K_{n\beta }$    
can
be dropped (since $K$ has no zero column). We find that  eq.   
\eqr{10} in   
this
case implies an equation for the matrix $A$, namely     
\begin{equation}
\sum_{\alpha=-\infty}^\infty H_{n \alpha }A_{\alpha m}=    0,   
\label{14}
\end{equation}
or in matrix form $HA=    0$ (however, the matrices are infinite).

Let us now turn to eq. \eqr{11} and make use of the finite   
range of the
interaction $F_{n}$ in eq. \eqr{1.1}. We have     
\begin{equation}
{\frac{\partial F_{n}}{\partial u_{k}}}=     
 0,\;\;n+n_{u}<k\;\;{\rm or}%
\;\;k<n-n_{d}  \label{15}
\end{equation}
for some non-negative integers $n_{u}$ and $n_{d}$.   
In eq. \eqr{11} all
exponentials, obtained after substituing for $F_{n}$   
from eq. \eqr{1.1}, are
linearly independent. This allows us to split eq.   
\eqr{11} into two   
types      
of
equations. These are obtained as coefficients of
 $\exp \left( {{%
\sum_{l}H_{ml}u_{l}}}\right) $, with $m\in
 [n-n_{1},n+n_{2}]$ and with $m$
outside this interval, respectively. Thus we have:     
\begin{equation}
\begin{array}{r}
\displaystyle{-K_{nm}\left[(\xi _{x}+\eta _{y})+\sum_{\alpha =       
m-n_{3}}^{m+n_{4}}B_{\alpha
} H_{m\alpha }\right]+
\sum_{\rho =    m-n_{1}}^{m+n_{2}}A_{n\rho }K_{\rho
m}=    0},\\ \\ \displaystyle{m\in [n-n_{1},n+n_{2}],}
  \label{16}
\end{array}
\end{equation}
\begin{equation}
\sum_{\rho =    m-n_{1}}^{m+n_{2}}A_{n\rho }K_{\rho m}=    
  0,\;\;m\not\!
{\in}[n-n_{1},n+n_{2}].  \label{17}
\end{equation}
We shall show that eq. \eqr{17} actually holds for all values   
of $m$ so that
eq. \eqr{16} can be simplified. To do this, we view  eq.
 \eqr{14} as a
difference equation for $A_{\alpha m}$. To make this   
explicit we   
restrict $H$    
and $K$ to be band matrices, with finite bands of constant width    
\begin{equation}
H_{n m}=    H_{n,\, n+\sigma }=    \left\{     
\begin{array}{l}
{h_{\sigma }(n)\;\;\sigma \in [p_{1},p_{2}]} \\     
{0\qquad\; \sigma \not\!{\in}\, [p_{1},p_{2}]}
\end{array}
\right. {,\; {h_{p_{1}}(n)\neq 0,\;\;h_{p_{2}}(n)\neq 0}.}    
 \label{18}
\end{equation}
Similarly     
\begin{equation}
K_{n m}=    K_{m+\sigma, m}=    \left\{     
\begin{array}{l}
{k_{\sigma }(m)\;\;\sigma \in [q_{1},q_{2}]} \\     
{0\;\;\;\;\;\;\;\;\;\sigma \not\!{\in}\, [q_{1},q_{2}]}
\end{array}
\right. {,\; k_{q_{1}}(m)\neq 0,\;\;k_{q_{2}}(m)\neq 0.}     
\label{19}
\end{equation}
In these notations we see that eq. \eqr{14} is a linear
 difference      
equation
for $A_{\sigma m}$ with $p_{1}-p_{2}+1$ terms     
\begin{equation}
\sum_{\sigma =    p_{1}}^{p_{2}}h_{\sigma }(n)A_{\sigma +n,
 m}=    0.      
\label{14'}
\end{equation}
Equation  \eqr{14'} determines the dependence of $A_{nm}$
 on $n$. Indeed   
the
linear difference equation     
\begin{equation}
\sum_{\sigma =    p_{1}}^{p_{2}}h_{\sigma }(n)\   
\psi _{\sigma +n} =    0
\label{diff-1}
\end{equation}
has $p_{2}-p_{1}$ linearly independent solutions, a
basis of which we denote    
by $\left\{ \psi _{n}^{j},\;j=    1,2,\dots ,
p_{2}-p_{1}\right\} $.    
Thus,  we      
have     
\begin{equation}
A_{n m}=    \sum_{j=    1}^{p_{2}-p_{1}}\psi _{n}^{j}C_{jm} ,   
\label{semi-A}
\end{equation}
where $C_{j m}$ are constants to be determined by the remaining      
determining
equations \eqr{16} and \eqr{17}. In order to analyze them,   
let us define   
the
quantities     
\[
Q_{n m}=    \sum_{\sigma =    m-n_{1}}^{m+n_{2}}
A_{n\sigma }K_{\sigma
m}.     
\]
From eq. \eqr{17} we have $Q_{n m}=    0$ for $m$   
``sufficiently far     
away''      
from $
n $. But, by using the expansion \eqr{semi-A}, we get     
\[
{Q_{nm}=    \sum_{j=    1}^{p_{2}-p_{1}}\psi _{n}^{j}\sum_{\sigma
 =     m-n_{1}}^{m+n_{2}}C_{j\sigma }K_{\sigma m}}     
\]
which, because of the linear independency of the $\psi _{n}^{j}$,      
implies     
\begin{equation}
\sum_{\sigma =    m-n_{1}}^{m+n_{2}}C_{j\sigma }K_{\sigma m}=    0    
\label{diff-C}
\end{equation}
for all values of $m$, since this relation does not depend
 on $n$ and      
the
index $m$ is no longer tied to $n$. In other words, if   
$Q_{nm}=    0$      
holds for
certain values of $n$ and $m$, as in eq. \eqr{17}, then   
that equation      
must
hold for all values. As in the case of      
eq.       
\eqr{14'}, we introduce a solution basis $\left\{ \phi_{m}^{l},l=        
1,\ldots
,q_{2}-q_{1}\right\} $ for the equation     
\begin{equation}
\sum_{\sigma =    q_{1}}^{q_{2}}k_{\sigma }\left( m \right) \ \phi   
_{\sigma      
+m}=    0.
\label{diff-2}
\end{equation}
 The general solution of eq. \eqr{diff-C} now is     
\[
C_{j m}=       
\sum_{l=1}^{q_{2}-q_{1}} q_{j l}\; \phi_{m}^{l},     
\]
where $q_{j l}$ are arbitrary constants. The expression   
\eqr{semi-A}      
for $A_{n m}$ is replaced by     
\begin{equation}
A_{n m}=    \sum_{j =    1}^{p_{2}-p_{1}}
{\sum_{l =   1}^{q_{2}-q_{1}} }
q_{j l}\;{\psi _{n}^{j}} \phi _{m}^{l}.  \label{A}
\end{equation}
A further consequence is that the last term in eq.  \eqr{16} can be      
dropped.
Then,  using the general solution for eq.  \eqr{9}      
\[
B_{n}\left( x,y\right) =    \beta _{n}\left( x\right) +
\gamma _{n}\left(
y\right),    \]
we separate the $x$ from the $y$ dependence in eq. \eqr{16}
 and  reduce      
  it  to two inhomogeneous difference equations for
 $ \beta _{n}\left( x\right)$ and $\gamma _{n}\left(
y\right)$. The general solutions of which are
\begin{equation}
{\beta _{n}}\left( x\right) =    
 \stackrel{p_{2}-p_{1}}{{\sum_{j=    1}}}{r_{j}}
\left( x\right)  \psi _{n}^{j} - b_{n} \xi_{x}(x),\qquad   
\gamma _{n}
\left( x\right) {=    }\stackrel{p_{2}-p_{1}}{{\sum_{j=        
1}}}s{_{j}}\left(
y\right) \ \psi _{n}^{j} -b_{n} \eta_{y}(y),  \label{beta-sol}
\end{equation}
where $b_{n}$ is an arbitrarily chosen solution of the inhomogeneous
 difference equation
\begin{equation}
\sum_{\sigma =    p_{1}}^{p_{2}}h_{\sigma }(n)\ b_{\sigma +n} =        
1.
\label{diff-3}
\end{equation}
Furthermore, in eq. \eqr{beta-sol}  the functions   
$r_{j}\left( x\right)     
$ and    
$s_{j}\left( y\right) $ are
arbitrarily chosen.
Finally, we obtain the following theorem.
\begin{theorem}
Consider all the generalized Toda theories of the form \eqr{1.1} for
infinitely many fields $u_{n}\left( x,y\right) $, where the coupling
matrices $H$ and $K$ satisfy eqs. \eqr{18} and \eqr{19}. Their Lie      
point
symmetry algebra is infinite-dimensional and a basis for it   
is given by      
  the following vector fields:     
\begin{equation}
{\hat{X}(\xi )}=    \xi (x)\partial_{x} -    
\xi_{x}\left(x\right) \sum_{n=-\infty}^\infty b_{n} \partial_{u_{n}},\quad    
\hat{Y}(\eta )=    \eta (y)\partial_{y} -    
\eta_{y}\left(y\right)\sum_{n=-\infty}^\infty b_{n} 
\partial _{u_{n}}, \label{conf}
\end{equation}
\begin{equation}
\hat{U}_{j}\left( r_j\right) =    r_{j}\left( x\right)   
\sum_{n=-\infty}^\infty\psi      
_{n}^{j}\partial
_{u_{n}},\;\;\hat{V}_{j}\left( s_j\right) =    s_j\left( y\right)
 \sum_{n=-\infty}^\infty\psi
_{n}^{j}\partial _{u_{n}}\;\;\left( j=    1,\dots ,p_{2}-
p_{1}\right) ,
\label{gauge}
\end{equation}
\begin{equation}
\hat{Z}_{jl}=    \left( \sum_{m=-\infty}^\infty\phi _{m}^{l}u_{m}\right)   
\left(      
\sum_{n=-\infty}^\infty\psi
_{n}^{j}\partial _{u_{n}}\right) \quad \left( j=    1,\dots
,p_{2}-p_{1};\,l=    1,\ldots ,q_{2}-q_{1}\right) .  \label{Z}
\end{equation}
The functions $\xi (x),\,\eta (y),\, r_j\left( x\right) $ and   
$s_j\left(      
y\right) $
are arbitrary, all the other quantities are determined by   
solving the      
linear
difference eqs.  \eqr{diff-1}, \eqr{diff-2} and \eqr{diff-3}.
\end{theorem}
As far as interpretation is concerned, we see that the generalized      
$\infty -$
Toda lattice \eqr{1.1} is always conformally invariant, since   
the vector
fields \eqr{conf} generate arbitrary reparametrizations of $x$   
and $y$,
accompanied by appropriate transformations of the fields   
$u_{n}$.
More specifically, the conformal transformations leaving eq.   
\eqr{1.1}     
invariant
are
\begin{equation}
\begin{array}{l}
\displaystyle{
\tilde{x}=  \psi(x,\lambda),\quad\tilde{y}=  \chi(y,\lambda),}\\ \\
\displaystyle{\tilde{u}_n(\tilde{x},\tilde{y})=  u_n(x,y)-b_n
\ln{(\frac{{    
\mbox    
d}\psi}{{\mbox d}x}\,
 \frac{{\mbox d}\chi}{{\mbox d}y})},}
 \end{array}
 \end{equation}
 where $\psi(x,\lambda)$ and $\chi(y,\lambda)$ are arbitrary   
functions
 of $x$ and $y$, related to $\xi(x)$ and $\eta(y)$ by the relations
 \begin{equation}
 \begin{array}{l}
 \displaystyle{
 \tilde{x}=  \psi(x,\lambda)=  T^{-1}(\lambda+T(x)),}\\ \\
 \displaystyle{\tilde{y}=  \chi(y,\lambda)=  S^{-1}(\lambda+S(y)),}
 \end{array}
 \end{equation}
 with
 \begin{equation}
 T(x)=  \int_{0}^x{\frac{{\mbox d}s}{\xi(s)}},
 \qquad\qquad
 S(y)=  \int_{0}^y{\frac{{\mbox d}t}{\eta(t)}}.
 \end{equation}
 The      
vector
fields $\hat{U}_{j}\left( r\right) $ and $\hat{V}_{j}
\left( s\right) $
generate gauge transformations: certain functions obtained by      
integrating
the vector fields can be added to any solution. Formally,   
the operators      
$\hat{Z}_{j l}$ generate linear transformations among   
components of      
solutions.
However, the sums are over infinite range, so convergence   
problems may
arise. Moreover, we have     
\begin{equation}
\partial _{xy}\left( \sum_{m}\phi _{m}^{l}u_{m}\right) =    0     
\end{equation}
as a consequence of eq.  \eqr{diff-2}. In other words,   
if the equation \eqr{diff-2}  admits non trivial solutions, than one   
can always perform a linear transformation among the $u_{n}$'s,
 in such a way
$q_{2} - q_{1}$ new fields $v_{l} = \sum_{m} \phi^{l}_{m} u_{m}$,
satifying the wave equation $\de_{x} \de_{y} v_{l} = 0$, are replaced
in the Toda system.   

As stated in Theorem 1, the problem of finding all   
symmetries of eq.      
\eqr{1.1}    
reduces to solving the  recursion relations \eqr{diff-1},   
\eqr{diff-2}      
 and \eqr{diff-3}. In general, this  may not be possible      
analytically in closed form.  Well developed      
techniques
exist for solving homogeneous and inhomogeneous difference   
equations      
with
constant coefficients \cite{32,33}. This is the case that   
occurs for all
generalized Toda field theories that we found in the literature:      
$h_{\sigma
}\left( n\right) $ and $k_{\sigma }\left( m\right) $ do not   
depend on      
$n$ and $m$, respectively.
\smallskip The nonzero commutation relations for the   
symmetry algebra     
of      
the
generalized $\infty -$Toda theory \eqr{1.1} are:     
\begin{equation}
\begin{array}{rl}
\ds{
\left[\hat{X}(\xi_1),\,\hat{X}(\xi_2)\right] =    \hat{X}(\xi      
_1\;\xi_{2, x} - \xi_{1, x}\;\xi_{2}),}
&\ds{\left[\hat{Y}(\eta      
_{1}),\hat{Y}(\eta
_{2})\right] =    {\hat{Y}({\eta}_{1}\;
{\eta }_{2, y}-{\eta }_{1,
y}\;{\eta      
}_{2}),}}
 \\ \\
 \ds{
\left[ {\hat{X}(\xi ),}\;\hat{U}_{j}\left( r\right) \right] =        
\hat{U}%
_{j}\left(\xi r_x \right),} &\ds{\left[ {\hat{Y}(\eta ),}
\;\hat{V}_{j}\left(
s\right) \right] =    \hat{V}_{j}\left( \eta \ s_{y}\right),}
\\ \\
\ds{\left[ {\hat{X}(\xi ),}\;\hat{Z}_{jl}\right] =     
 -\hat{U}_{j}\left(
\xi_x\sum_n b_n\phi_n^l\right),} &\ds{\left[ {\hat{Y}(\eta      
),}\;\hat{Z}_{jl}\right] =    %
-\hat{V}_{j}\left( \eta_y\sum_n b_n\phi_n^l\right),}    
\\ \\
\ds{\left[ \hat{U}_{a}\left( r\right) ,\hat{Z}_{jl}\right]      
=    \hat{U}_{j}\left(
r\sum_{m}\phi _{m}^{l}\psi _{m}^{a}\right),} &\ds{\left[     
\hat{V}_{a}\left(
s \right) ,\hat{Z}_{jl}\right] =    
  \hat{V}_{j}\left( s\sum_{m}\phi      
_{m}^{l}\psi
_{m}^{a}\right),}
\\ \\
\ds{\left[ \hat{Z}_{ab},\hat{Z}_{cd}\right] =    
  \left( \sum_{m}\phi      
_{m}^{d}\psi
_{m}^{a}\right) \hat{Z}_{c b}-}&\ds
{\hspace{-3mm}\left( \sum_{m}\phi     
_{m}^{b}    
\psi_{m}^{c}\right) \hat{Z}_{ad}.} \label{commutrel}
\end{array}
\end{equation}
The algebra of vector fields \smallskip $\hat{Z}_{jl}$ is finite      
dimensional
(its dimension is $d=    \left( p_{2}-p_{1}\right) \times \left(
q_{2}-q_{1}\right) $). However, its isomorphism class cannot be      
determined
without specifying the functions $\phi _{m}^{l}$   
and $\psi _{n}^{j}$,      
i.e.
the matrices $H$ and $K$ in \eqr{1.1}. In all examples in the     
literature,      
we
have either $d=    1$, or $d=    0$. It is however easy to invent     
examples      
in which $\left\{ \hat{Z}_{jl}\right\} $ is simple,   
semisimple, solvable, or    
whatever
we postulate a priori.

The overall structure of the obtained Lie algebra is     
\begin{equation}
\left( \left\{ {\hat{X}}\right\} \oplus  \left\{ {\hat{Y}}\right\}      
\right)
\raisebox{1pt}{$\scriptstyle +$}\hspace{-3.7mm}\supset
 \left( \left\{ \hat{Z}\right\} \raisebox{1pt}{$\scriptstyle    
+$}\hspace{-3.7mm}\supset \left( \hat{U}\oplus      
\hat{V} \right) \right) .  \label{alg-strutt}
\end{equation}

If $\left\{ \hat{Z}\right\} $ is solvable, then   
\eqr{alg-strutt} amounts to
a Levi decomposition, since both $\left\{ {\hat{X}}\right\} $ and      
$\left\{ {%
\hat{Y}}\right\} $ are centerless Virasoro algebras and   
hence simple. We
recall that the Levi theorem does not hold for   
infinite-dimensional Lie
algebras and a Levi decomposition does not necessarily exist.

Let us sum up the general results obtained so far for the
 symmetries of      
the
generalized $\infty -$Toda field theories \eqr{1.1} under the     
constraints
imposed in Theorem 1.
\begin{enumerate}
\item   The theory is always conformally invariant, since the      
inhomogeneous
equation \eqr{diff-3} always has a solution.
\item   The theory allows gauge transformations $\hat{U}$ and      
$\hat{V}$ if $p_{2}-p_{1}\geq 1$.
\item   The transformations of type $\hat{Z}$ exist if $\left(
p_{2}-p_{1}\right) \left( q_{2}-q_{1}\right) \geq 1.$
\end{enumerate}

\subsection{\protect\bigskip \noindent {\sl Special cases}      
\protect\medskip
\noindent}

\noindent{\bf {1.} The Mikhailov-Fordy-Gibbons  two   
dimensional $\infty$-Toda    
system \eqr{1.2}}

We have     
\begin{equation}
h_{-1}\left( n\right) =    -h_{0}\left( n\right) =    1,\quad   
{\rm and\;}     
k_{-1}\left(n\right) = -k_{0}\left( n\right) =
    -1 ,\label{MFGcoef}\end{equation}  so
$p_{2}-p_{1}=    q_{2}-q_{1}=    1$.    
From eqs. \eqr{diff-1} and \eqr{diff-2} we have
\[
\psi _{m}=    \phi _{m}=    1.
\]
 Equations \eqr{beta-sol} and \eqr{diff-3} in this case imply
\[
\beta _{n}=    \beta (x)+n\xi _{x},\;\;\gamma _{n}=    
  \gamma (y)+n\eta_{y}.
\]
From Theorem 1 we now obtain all symmetries of eq \eqr{1.2}, namely
\begin{eqnarray}
{\hat{X}(\xi )=    \xi (x)\partial _{x}+\xi_{x}\sum_{n=-\infty}^\infty n 
\partial_{u_{n}},\; \hat{Y}(\eta )} &=    &{\eta(y)\partial_{y}+
\eta_{y}\sum_{n=-\infty}^\infty n\partial_{u_{n}}}, \nonumber\\
\hat{U}=    \beta (x)
\sum_{n=-\infty}^\infty\partial_{u_{n}},\qquad\hat{V}&=     
&     
\gamma
(y)\sum_{n=-\infty}^\infty\partial _{u_{n}}, \label{symmMFG} \\
\hat{Z}=    \left( \sum_{m=-\infty}^\infty u_{m}\right) \left(   
\sum_{n=-\infty}^\infty\partial_{u_{n}}\right).     
&&\nonumber   
\end{eqnarray}
The generators ${\hat{X}}, {\hat{Y}},
{\hat{U}}$ and $\hat{V}$ were obtained       
in
ref.\cite{21} using the so called ``intrinsic method''.
 The generator      
$\hat{Z}$ was not obtained there   
and cannot be obtained by the intrinsic    method.

\noindent{\bf {2.} The Toda field   
theory \eqr{1.3}}

We take $H=    I$. Then equations \eqr{diff-1}, \eqr{diff-2} and     
\eqr{diff-3}    
in this case imply     
\[
{\beta _{m}=    -\xi _{x},\;\;,\gamma _{m}=    -\eta _{y},\quad     
A_{nm}=    0.}
\]
The theory is only conformally invariant    
\begin{equation}
{\ \hat{X}(\xi )=    \xi (x)\partial _{x}-\xi _{x}\sum_{n}\partial     
_{u_{n}},\quad\hat{
Y}(\eta )=    \eta (y)\partial _{y}-\eta _{y}\sum_{n}
\partial _{u_{n}}}
\label{LezSavsymm}\end{equation}
and no further symmetries are obtained.

\noindent{\bf {3.} The Toda field theories \eqr{1.4}}

 We take $K=I$ and relation \eqr{diff-2}
implies     
\[
A_{nm}=    0.
\]
The remaining equations \eqr{diff-3} cannot be solved   
explicitely for
general $h_{\sigma }(m)$, but as said above, we can   
easily deal with in      
the
constant coefficients case. As an example,   
let us restrict to the case when $H$  is the $A_{\infty}$   
 Cartan matrix (This is the $A_{N}$
 Cartan
matrix for $N \rightarrow \infty $, where the limit   
is taken symmetrically from a fixed, but not extremal,   
vertex in the corresponding Dynkin diagram). Thus we have     
\begin{equation}
h_{-1}=    h_{+1}=    -1,\;\;h_{0}=    2, \label{BilalGer}
\end{equation}
the solutions \eqr{beta-sol} become     
\begin{equation}
\ \beta_{n}=    \frac{n^{2}}{2} \, \xi_{x}+n \,
r_{2}(x)+r_{1}(x),\quad     
\gamma_{n}=    \frac{n^{2}}{2} \, \eta_{y}+n\, s_{2}(y)+s_{1}(y).
\label{solBilGer} \end{equation}
The symmetry algebra is     
\begin{eqnarray}
\hat{X}(\xi ) &=    &\xi (x)\partial_{x} + \frac{1}{2} \xi     
_{x}\sum_{n=-\infty}^\infty n^{2} \partial_{u_{n}},\quad    
\hat{Y}(\eta ) =    \eta (y)\partial_{y} + \frac{1}{2} \eta
_{y}\sum_{n=-\infty}^\infty {n}^{2}\partial _{u_{n}}, \nonumber \\
\hat{U}_{1}\left( r_1\right)  &=    &r_{1}\left( x\right)
 \sum_{n=-\infty}^\infty\partial
_{u_{n}},\qquad \hat{V}_{1}\left( s_1\right) =     
 s_{1}\left( y\right)    
\sum_{n=-\infty}^\infty\partial_{u_{n}}, \label{symmBilGer}\\
\hat{U}_{2}\left( r_2\right)  &=    &r_{2}\left( x\right)
 \sum_{n=-\infty}^\infty n\,
\partial_{u_{n}},\qquad \hat{V}_{2}\left( s_2\right) =     
 s_{2}\left(     
y\right) {\sum_{n=-\infty}^\infty n\partial _{u_{n}},} \nonumber   
\end{eqnarray}
where $\xi (x)$, $\eta (y)$, $r_{1}(x)$, $s_{1}(y)$,   
$r_{2}(x)$ and     
$s_{2}(y)$ are arbitrary smooth functions.

\section{{\ Symmetries of Finite Generalized Toda Field Theories }}
\subsection{\protect\bigskip \noindent {\sl General Results}      
\protect\medskip
\noindent}
In this case we have a system of $N$ partial differential equations
 in $N$ fields $u_{n}\lf(x , y \ri)$, namely
\begin{equation}
 u_{n, x y} =   F_{n},  \qquad    
F_{n} =    
\sum_{m =   1}^{M} K_{n m}
\exp\lf( \sum_{l =   1}^{N} H_{m l} u_{l}\ri) \quad \lf(1 \leq n \leq N\ri) .     
\label{f1}
\end{equation}
The ``coupling constant''  matrices $H$ and $K$ satisfy  $H \in
{\mathbb R}^{M \times N}$ and $K \in {\mathbb R}^{N \times M}$.    
The system \eqr{f1} could  arise in a quite general    
field theory with  Lagrangian
\begin{equation}
{\cal L} =   \frac{1}{2} \sum_{m , n  =   1}^{N} \kappa_{m n}    
\de_{x} u_{m} \de_{y} u_{n} -
\sum_{m =   1}^{M} c_{m} \exp \lf
(  \sum_{l =   1}^{N} H_{m l} u_{l}\ri) \qquad \lf( c_{m}   
\neq 0 \ri),
\label{f2}\end{equation} with
\begin{equation}
K =   L^{-1} H^{T} C, \qquad L =   \frac{ \kappa+\kappa^{T}}{2},
 \qquad  C =   {\rm diag} \lf( c_{1}, \dots , c_{N} \ri) .   
\label{f3}
\end{equation}
Some general considerations concerning the system \eqr{f1}   
are in order.     

First, if either  $K$, or $H$ (or both) allow an inverse,
 or at least a   
left inverse, then this system can be simplified. Indeed,   
let $K^{-1}$     
exist. We put    
$u_{n}=   \sum_{m} K_{n m} \rho_{m}$ and obtain    
\begin{equation}
\rho_{m, x y} =   \exp \lf( \sum_{l=  1}^{M}\lf( H K \ri)_{m l}     
\rho_{l}\ri ), \quad 1\leq m \leq M .\label{TodatrBG}
\end{equation}
Conversely, let $H^{-1}$ exist and put $w_{j} =   \sum_{l} H_{j l}     
u_{l}$, we obtain    
\begin{equation}
w_{m, x y}   =    \sum_{j=  1}^{M}\lf( H K \ri)_{m j} e^{w_{j}},
 \quad     
1\leq m \leq M .\label{TodatrLS}\end{equation}
In other words, one of the matrices    
$H$, or $K$ can be normalized to $I_{M}$, if it is left invertible.

The second comment is that the system \eqr{f1} with $K =    
 I$ admits Lie-B\"acklund transformations, and in this sense
 is completely  integrable, if the matrix $H$ is a Cartan, 
or a generalized Cartan matrix     
\cite{19}.

We  mention that in the case of the infinite Toda field
theories the     
matrices $H$ and $K$ in general have nontrivial kernels,   
are hence not     
invertible and we cannot normalize them.

Let us now turn to the Lie point symmetries of the system   
\eqr{f1}.
 We write a general element of the symmetry algebra in the   
form \eqr{4}
 (with the sum in the range $1 \leq n \leq N$),    
apply its prolongation to eq. \eqr{f1} as in eq. \eqr{6}.    
From the determining equations we find that for any $F_{n}$    
in eq. \eqr{f1}, in complete analogy with the    
$\infty$-Toda theory,  a general element of the symmetry    
algebra will have the form  \eqr{7}, the summation being from $1$ to $N$.
   
Two determining equations remain and they   
depend on the     
specific form of $F_{n}$ in eq. \eqr{f1}. Making use of   
the fact that     
all the exponentials are linearly independent (no two   
rows in $H$     
coincide) and that the matrix $K$ has no zero column,   
we reduce the     
remaining determining equations to two matrix relations
\begin{equation}
HA =   0 , \label{findeteq1}\end{equation}
\begin{equation}
\lf[ \lf ( A - \lf( \xi_{x} +\eta_{y}\ri )I\ri)    
K \ri]_{n m} =   K_{n m} \lf( H B \ri)_{m}\;\;    
\lf(1 \leq n \leq N ,\;\; 1 \leq m \leq M\ri).\label{findeteq2}\end{equation}
We multiply eq. \eqr{findeteq2} by $H$ from the left and use    
\eqr{findeteq1} to obtain
\begin{equation}
- \lf(\xi_{x}+\eta_{y} \ri)\lf( H \, K \ri)_{k m}  =      
  \lf( H \, K \ri)_{k m} \lf(H B \ri)_{m} \;\; \forall k,m .
 \label{constr1} \end{equation} If the matrix $HK$ has no   
zero column, then we obtain \begin{equation}
HB=  -\left( \xi _{x}+\eta _{y}\right)
{\bf {\bar{1}}}_{M},
 \label{HB}\end{equation}
 where ${\bf {\bar{1}}}_{M}=  \left(1,\dots,1\right)^T\in {\mathbb R}^{M}$,
and from eq. \eqr{findeteq2}    
\begin{equation}
AK =   0. \label{AK}
\end{equation}
Thus, matrix $A$ must satisfy the same two    
homogeneous equations \eqr{findeteq1} and \eqr{AK}
 as in the infinite case. Furthermore, if ${\bf {\bar{1}}}_{M}$
 is in
 the image of $H$, then we define ${\bf b}_{N}  \in {\mathbb R}^{N}$ to be
an
 arbitrarily chosen (but specified) solution of the   
inhomogeneous    
equation    
\begin{equation}
H{\bf b}_{N} =   {\bf {\bar{1}}}_{M}. \label{Hk}
\end{equation}
The results of these considerations can be summed up as follows    
\begin{theorem}
 Consider the generalized Toda field theories \eqr{f1}
with a
finite number of fields $N$. Assume that all rows in $H$
 are different and
that the matrix $H\,K$ has no zero column. Then 3 types    
of symmetries can
occur and they depend on the properties of the    
fundamental spaces of the
matrices $H$ and $K$.
The symmetries are of the same form as in Theorem 1, except    
that all    
summations range from 1 to $N$.
 However, if   ${\bf {\bar{1}}}_{M} \in Im(H)$,
 then $\xi$ and $\eta$ are arbitrary functions of
 $x$ and $y$, respectively, and the theory is conformally   
invariant.    
The quantities $b_{n} $ are the    
components of the vector ${\bf b}_{N}$, itself an arbitrary    
solution of eq. \eqr{Hk}. Otherwise,    
if ${\bf {\bar{1}}}_{M} \not \in Im(H)$,    
the theory is invariant    
only under the Poincar\'e group, generated by    
\begin{equation}
    {\hat P}_{1} =   \partial_{x}  ,\quad    
 {\hat P}_{2} =   \partial_{y} ,\quad    
{\hat  L} =   x \partial_{x} - y \partial_{y}.
\label{Poincare}\end{equation}
Gauge transformations exist only if $H$ is not invertible.
 Analogously  to the formulas  \eqr{gauge},  $r_{j}$ and $s_{j}$    
are arbitrary    
functions and the vectors ${{\bf \psi}}^{j}$ span $Ker
\left( H\right)$.    
Finally, the vectors ${\bf \phi}^{l}$ span the left kernel   
of $K$.
If this space is not zero, then ${\rm dim}
\lf(Ker\lf( K^{T}\ri)\ri)   
\times {\rm dim}\lf( Ker\lf( H \ri)\ri)$   
symmetries of the form \eqr{Z}    
are admitted. \end{theorem}                        
From Theorem 2, contrary to the case of infinitely many fields,    
conformal    
invariance is not a priori  guaranteed, but    
it imposes restrictions on the    
image of $H$. Gauge symmetries exist only if    
the matrix $H$ has a nonzero    
kernel.

\subsection{\protect\bigskip \noindent {\sl Special cases}      
\protect\medskip
\noindent}

\noindent{\bf {1.} The Mikhailov-Fordy-Gibbons Toda theory    
and generalizations}

Consider the field equation    
\begin{equation}
{\bf U}_{x y}  = \frac{\mu^2}{\beta}    
\sum_{i =   1}^{N}\frac{\alpha_{i}}{\alpha_{i}^2}
\exp\lf( \beta \alpha_{i} \cdot {\bf U}\ri), \label{MFG}
\end{equation}
where ${\bf U} =  \lf( u_{1}, \dots, u_{N}\ri) $
is an N-ple of real fields and $\lf({\bf \alpha}_{1},\dots,
{\bf \alpha}_{N} \ri)$ denote
the simple roots of a classical simple finite Lie algebra.
Equations \eqr{MFG} above take the form \eqr{1.2} for all $n$ satisfying
 $N_{0}\leq  n\leq  N-1$. For $n=N$ we obtain
\begin{equation}
u_{N,xy}=  \exp\lf(u_{N-1} -u_{N}\ri).
\end{equation}
The equations for $1\leq n<N_0$ are different for each Cartan series.
The number  $N_{0}$ is equal to 2 for $A_{N}, B_{N}, C_{N}$, and 3    
for $D_{N}$.

For the $A_{N}$ algebra we have    
\begin{equation}
u_{1,x y} =   -\exp\lf(u_{1}-u_{2}\ri).
\label{1AN}\end{equation}
 Conformal and gauge transformations are exactly the    
same as given in eq. \eqr{symmMFG} (except   
that the summations are from $1$ to $N$).    

For the $ B_{N}$  algebra we have
\begin{equation}
u_{1, x y} =   \exp\lf(-u_{1}\ri) - \exp\lf(u_{1}-u_{2}\ri).
\label{1BN}
\end{equation}
Conformal transformations are as in eq.    
\eqr{symmMFG} (with the same comment about the summations)
 and there is no gauge invariance.

For the $C_{N}$ algebra we have
\begin{equation}
u_{1,x y} =   -\exp\lf(u_{1}-u_{2}\ri)+2 \exp\lf(-2 u_{1}\ri).
\label{1CN}\end{equation}
The only symmetry is conformal invariance,
generated by
\begin{eqnarray}
\hat{X}(\xi )& = &   \xi (x)\partial_{x} + \xi_{x}
\sum_{n=1}^N \lf ( n-\frac{1}{2} \ri) \partial_{u_{n}},\nn  \\
\hat{Y}(\eta )& =  &  \eta (y)\partial_{y} + \eta_{y}
\sum_{n=1}^N \lf ( n-\frac{1}{2} \ri) \partial_{u_{n}}
\label{MFGCN}\end{eqnarray}

Finally, for the $D_{N}$ algebra we have
\bea
u_{1, x y} &=  & \exp\lf(-u_{1} - u_{2}\ri) - \exp\lf(u_{1}
 - u_{2}\ri),\nn \\
u_{2, x y} &=  & \exp\lf(-u_{1}-u_{2}\ri) + \exp\lf(u_{1}-u_{2}\ri)
- \exp\lf(u_{2} - u_{3}\ri).
\label{1DN}\eea
Again,  the only symmetry is conformal invariance, in this case generated
by
\begin{eqnarray}
\hat{X}(\xi )& = &   \xi (x)\partial_{x} + \xi_{x}
\sum_{n=1}^N \lf ( n- 1 \ri) \partial_{u_{n}},\nn  \\
\hat{Y}(\eta )& =  &  \eta (y)\partial_{y} + \eta_{y}
\sum_{n=1}^N \lf ( n- 1 \ri) \partial_{u_{n}}.\label{confDN}
\end{eqnarray}

We mention that the infinite system \eqr{1.2} can also be
 reduced to the   
finite    
one by imposing periodicity $u_{N+1} =  u_{1}$.
 In this case ${\bf {\bar{1}}}_{N}$ is not contained   
in  $Im(H)$ and there is no conformal invariance. Thus, the   
symmetry is given  by the two dimensional Poincar\'e algebra   
\eqr{Poincare} and by the gauge generators given in \eqr{symmMFG}.

\noindent {\bf 2. The Toda field theory \eqr{1.3}}

 The symmetries are the same in the finite
 case as in the infinite one, namely the conformal transformations   
generated by   
\eqr{LezSavsymm} (for any finite matrix $k$).

\noindent {\bf 3. The finite Toda theories \eqr{1.4}}

 Since the Cartan matrix $H$ is   
invertible,
 this theory is equivalent to that described by eq. \eqr{1.3} in   
the sense of eqs. \eqr{TodatrBG} and \eqr{TodatrLS}.   
Hence this theory is always and only conformally invariant.   
However, the generators of the vector fields take a
 slightly different form, which we report for a subsequent   
discussion.

For the $A_{N}$ algebra the generators are given by
\begin{equation}
{\hat {W}} = \xi (x)\partial_{x} + \eta (y)\partial_{y} +   
\frac{1}{2} \lf( \xi_{x}+ \eta_{y}\ri)   
\sum_{n = 1}^{N} n \lf ( n - N - 1 \ri) \partial_{u_{n}}.
\end{equation}

For the $B_{N}$ algebra, the symmetry generator is given by
\begin{eqnarray}
&\ds{{\hat {W}} = \xi (x)\partial_{x} + \eta (y)\partial_{y} -
\frac{1}{4} \lf( \xi_{x}+ \eta_{y}\ri)\times} \\
&\ds{ \lf \{  N \lf( N+1\ri)
\partial_{u_{1}} +
2 \sum_{n = 2}^{N} \lf[ N \lf( N+1\ri) - n \lf ( n  - 1 \ri)\ri]
\partial_{u_{n}}\ri \}.}
\end{eqnarray}

For the $C_{N}$ algebra, the symmetry generator is given by
\begin{equation}
{\hat {W}} = \xi (x)\partial_{x} + \eta (y)\partial_{y}   
+\frac{1}{2}\lf( \xi_{x}+ \eta_{y}\ri) \sum_{n = 1}^{N} \lf[   
n \lf( n-2 \ri) - N^{2} +1\ri]
\partial_{u_{n}}.
\end{equation}

Finally, for the $D_{N}$ algebra ($N \geq 4$), one has
\begin{eqnarray}
&\ds{{\hat {W}} = \xi (x)\partial_{x} +   
\eta (y)\partial_{y} - \frac{1}{4} \lf( \xi_{x}+ \eta_{y}\ri)   
\times} \\ &\ds{\lf \{  N \lf( N-1\ri)\lf(
\partial_{u_{1}} +\partial_{u_{2}}\ri) +  2   
\sum_{n = 3}^{N} \lf[ N \lf( N-1\ri) - \lf( n - 2 \ri)   
\lf ( n  - 1 \ri)\ri]
\partial_{u_{n}}\ri \}.}
\end{eqnarray}

\section{{\ Symmetries of Generalized Semi-Infinite
 Toda Field Theories }}

\subsection{\protect\bigskip \noindent {\sl General Results}      
\protect\medskip
\noindent}

Let us now restrict the range of the discrete   
variable $n$ to be $1 \leq n < \infty$.   
Both the equations \eqr{1.1} of the   
generalized Toda field theories, and their   
symmetries will be modified. The matrices
 $H$ and $K$ will no longer be pure band matrices but will have the form
\begin{equation}
{\tiny
H=\left(   
\begin{array}{llllll}
H_{1,1} &   
\begin{array}{lll}
\ldots  & \ldots  & \ldots   
\end{array}
& H_{1,N} &  &  &   
\begin{array}{ll}
&   
\end{array}
\\   
\ldots  &   
\begin{array}{lll}
\ldots  & \ldots  & \ldots   
\end{array}
& \ldots  &  &  &   
\begin{array}{ll}
&   
\end{array}
\\   
H_{M,1}^{{}} &   
\begin{array}{lll}
\ldots  & \ldots  & \ldots   
\end{array}
& H_{M,N} &  &  &   
\begin{array}{ll}
&   
\end{array}
\\   
& H_{M+1,M+1+p_{1}} &   
\begin{array}{lll}
\ldots  & \ldots  & \ldots   
\end{array}
& \ldots  & H_{M+1,M+1+p_{2}} &   
\begin{array}{ll}
&   
\end{array}
\\   
&  & H_{M+2,M+2+p_{1}} & \ldots  &   
\begin{array}{llll}
\ldots  & \ldots  & \ldots  & \ldots   
\end{array}
&   
\begin{array}{ll}
H_{M+2,M+2+p_{2}} &   
\end{array}
\\   
&  &  & \ddots  &   
\begin{array}{llll}
\ddots  &  &  & \ddots   
\end{array}
&   
\begin{array}{lllll}
\ddots  &  &  &  & \ddots   
\end{array}
\end{array}
\right), \label{semiH}
}
\end{equation}
where $M+p_{1} \leq N \leq M + p_{2}$   
and the void entries are equal   
to zero.  Similarly, the matrix $K$ takes the  form
\begin{equation}
K=\left(   
\begin{array}{llllll}
K_{1,1} & \ldots  & K_{1,N^{\prime }} &  &  &  \\   
\ldots  & \ldots  & \ldots  & K_{N^{\prime }+1+
q_{1},N^{\prime }+1} &  &  \\   
K_{M^{\prime },1} & \ldots  & K_{M^{\prime },
N^{\prime }} & \ldots  &   
K_{N^{\prime }+2+q_{1},N^{\prime }+2} &  \\   
&  &  & \ldots  & \ldots  & \ddots  \\   
&  &  & K_{N^{\prime }+1+q_{2},N^{\prime }+1} &   
\ldots  & \ddots  \\   
&  &  &  & K_{N^{\prime }+2+q_{2},N^{\prime }+2} & \ddots  \\   
&  &  &  &  & \ddots   
\end{array}
\right) \label{semiK}
\end{equation}
where $N'+q_{1} \leq M' \leq N'+q_{2}$.
Although one could easily construct non trivial models,
 which do
 not fit
in the given scheme, they seem quite artificial and, moreover,
 all the cases which we found in the literature   
satisfy the   
above restrictions.
    
 We denote by $\tilde{H}$
and $\tilde{K}$ respectively,
the $M \times N$ and $M'   
\times N'$ matrices,  which can be   
extracted by taking the first $M$    
rows and the first $N$  columns from $H$   
and, in turn, the first $M'$  rows and the first $N'$
columns from $K$.

The symmetry algebra of the semi-infinite
Toda field theory equation can either be   
obtained directly, {\it ab initio}, or we can   
obtain it from the infinite case of Section 2,   
by adding appropriate boundary conditions and   
requiring that they be invariant.
As above, the functions $\xi\lf( x \ri)$,   
$\eta \lf( y \ri)$, $A_{m n}$ and
 $B_{n} \lf(  x , y \ri)$ must
 satisfy the remaining determining   
equations \eqr{9} - \eqr{11}. Following the same reasoning as   
in the finite case (see Section 3), we obtain the analogs
 of all the relations    
\eqr{findeteq1} - \eqr{AK}, where now all the labels and 
summations range from 1   
to $\infty$ (i.e. we   
take $N\rightarrow \infty$ in all formulas). 
The key equation of the discussion   
is eq. \eqr{HB}
 and its   
associated homogeneous system.
Here,  we separate the problem into  the finite subsystems
\begin{eqnarray}
{\tilde H} \; {\bf {\tilde B}}& =& 0, \label{HtildeBtildea}\\
{\tilde H} \; {\bf {\tilde B}}& =&  -
\left( \xi _{x}+\eta _{y}\right)
{\bf {\bar{1}}}_{M},   
\label{HtildeBtildeb}\end{eqnarray}
where ${\bf {\tilde B}} = \lf(B_{1}, \dots , B_{N} \ri)$,   
and a difference linear equation, which we can put again in
 the form
\eqr{diff-1}, or \eqr{diff-3} respectively,  for $n \geq M+1$.   
The eq. \eqr{HtildeBtildea} has   
$Ker \lf( {\tilde H}\ri)$   as its solution space.
 On the other hand, the difference equation   
\eqr{diff-1} has a $\lf( p_{2}-p_{1}\ri)$-dimensional
 solution space, the elements of which have the form
\begin{equation}
B_{n} = \sum_{j = 1}^{p_{2} - p_{1}} \alpha_{j} \psi_{n}^{j},
 \qquad   
n \geq M+1+p_{1},
\label{repr}\end{equation}
in terms of the basis $\lf \{ \psi_{n}^{j} \ri \}$.
Moreover, the difference eq. \eqr{diff-1} has only the zero
solution in the case $p_{1} = p_{2}$. But, because of the
 imposed restrictions   
on the form of $H$, in such a
 case the components of the vector ${\bf {\tilde B}}$   
are decoupled from the remaining $\lf( B_{N+1}, \dots \ri)$.   
This means
that the semi-infinite homogeneous  linear system $HB = 0$
has zero-dimensional kernel only if both the finite system 
\eqr{HtildeBtildea}
and the homogeneous difference eq. \eqr{diff-1} do.

Assuming now that $p_{1} < p_{2}$ and, moreover,   
that $M+p_{1}+1 \leq N$, the components
$\lf( B_{M+1+p_{1}}, \dots , B_{N}\ri)$ have to satisfy both   
the finite linear
eq. \eqr{HtildeBtildea} and the difference eq. \eqr{diff-1}.
Substituting the representation   
\eqr{repr} into \eqr{HtildeBtildea},   
we get $N - {\rm dim}\lf( Ker \lf( {\tilde H}\ri)\ri)$   
constraints on the   
$\lf \{ \alpha_{i}\ri \}_{i = 1, \dots , p_{2} - p_{1}}$. Thus,
if it results that
\begin{equation}
M-N+p_{2}+\dim \left( Ker\left( \tilde{H}\right) \right)
 =n_{0}>0,\label{ker}
\end{equation}
then the semi-infinite homogeneous system $HB = 0$ admits a   
$n_{0}$-dimensional kernel,   
spanned by the set of linearly   
independent functions $\lf \{ \chi_{n}^{j} \ri \}_{j = 1,
 \dots ,n_{0}}$.   

The above result implies that, if the constraint    
\eqr{ker} holds, then the semi-infinite
Toda model defined \eqr{semiH}   
and \eqr{semiK} possesses a symmetry   
group of gauge transformations, generated by
the $2 \times n_{0}$ vector fields
\begin{equation}
\hat{U}_{j}\left( r_j\right) =    
  r_{j}\left( x\right) \sum_{n=1}^{\infty}\chi_{n}^{j}
\partial_{u_{n}},\;\;\hat{V}_{j}\left( s_j\right)
= s_j\left( y\right) \sum_{n=1}^{\infty}\chi_{n}^{j}\partial   
_{u_{n}}\;\;\left( j=    1,\dots ,p_{2}-p_{1}\right).
\label{semigauge}
\end{equation}
As in the finite case, a semi-infinite  theory is
conformally invariant if the inhomogeneous
 eq. \eqr{HB} ( for semi-infinite matrices) has a solution.
 Thus, now  we must require that the vector
${\bf {\bar{1}}} = \lf( 1, 1, \dots \ri)$ be contained in
 $Im \lf( H \ri)$. But, as outlined above, the problem is   
reduced to finding a solution of the eq.    
\eqr{HtildeBtildeb} and of the difference eq.   
\eqr{diff-3}. The former equation   
is solved if   
\begin{equation}
{\bf {\bar{1}}}_{M}  \in Im\lf({\tilde H}\ri).
\label{HtildeB}
\end{equation}
 For the difference eq. \eqr{diff-3}
a solution always exists as seen in Sec. 2.
 Hence the structure of the matrix $H$ shown in \eqr{semiH} garantees that a   
solution of the total inhomogeneous system always exists, 
once eq. \eqr{HtildeB}   
is satisfied ici. In conclusion,   
the condition
\eqr{HtildeB} is not only necessary,   
but also sufficient to ensure the
conformal invariance of the given Toda theories.   

Finally, an analysis similar to the study of the gauge   
invariance can be performed for the $\hat Z$-type transformations,
which exist if a common solution of the two semi-infinite
homogeneous systems   
\begin{equation}
HA = 0, \qquad AK = 0   
\end{equation}
can be found. Thus, we are lead to the following theorem
\begin{theorem}
Consider the semi-infinite Toda field theory \eqr{1.1}, with
$H$ and $K$ given by \eqr{semiH} and \eqr{semiK}, respectively, and
with all rows of $H$  different. Moreover,
let $HK$ have no zero columns.
Then, the symmetry algebra depends on the   
fundamental spaces of the finite dimensional submatrices $\tilde H$
and $\tilde K$, on the solutions of the difference eqs.
 \eqr{diff-1} and \eqr{diff-3} for $n \geq M+1$ and, finally,
on the solutions of the difference eq. \eqr{diff-2} for
$m \geq N' +1$.   

The theory is conformally invariant if the condition   
\eqr{HtildeB} holds. The corresponding generators   
take the form \eqr{conf}. Otherwise,   
if \eqr{HtildeB} does not hold, the symmetry reduces to the   
Poincar\'e
group generated by \eqr{Poincare}.

A gauge transformation group, involving $2n_{0}$   
arbitrary functions of one variable, exists if   
the relation \eqr{ker} holds. The algebra generators take the form
\eqr{semigauge}. Finally, $\hat Z$-type gauge transformations
exist if not only \eqr{ker} holds, but also the
supplementary condition
\begin{equation}
N' - M' + q_{2} + {\rm dim}\lf(Ker \lf( {\tilde K}^{T} \ri)\ri) =
m_{0} > 0
\end{equation}
is satisfied. In such a case they form a   
Lie algebra of dimension $m_{0} \times n_{0}$.   
\end{theorem}

\subsection{\protect\bigskip \noindent {\sl Special cases}      
\protect\medskip
\noindent}
 Now let us consider the same three examples as in the previous
 Sections.

\noindent {\bf 1. Mikhailov-Fordy-Gibbons field theories}

 All  examples  of Section   
3.2 can be generalized to the semi-infinite case, simply allowing $N$ to go
 to   
$\infty$ for each   
classical Lie algebra. The equations labeled by $1 \leq n \leq N_{0}$
are explicitly given by \eqr{1AN}, \eqr{1BN}, \eqr{1CN}   
and \eqr{1DN}, respectively. Moreover, for $i \geq N_{0}$ 
the equations are the
same as in the infinite case, i.e. eq. \eqr{1.2}.   

For the $A_{ \infty +}$ algebra (We use this notation
 in order to   
distinguish this semi-infinite model from the previously
 introduced $A_{ \infty}$ infinite one)
we have $M = N = M' = N' = 0$ and hence the symmetries   
are exactly the same as in the infinite and in the
 finite cases (see eq.
\eqr{symmMFG}), where the summations are over the appropriate range.

For the $B_{\infty}$ algebra one has ${\tilde H} = - {\tilde K}
= \lf( -1 \ri)$, then also $M = N = M' = N' = 1$, as one can see
from \eqr{1BN}. Theorem 3 allows
 to establish that there are no gauge transformations of any
 kind and the generators of the conformal transformations   
are the same as given in \eqr{symmMFG}.

From eq. \eqr{1CN} one sees that   
${\tilde H} = - {\tilde K}
= \lf( -2 \ri)$
for the $C_{\infty}$ algebra, then $M = N = M' = N' = 1$.
Thus, Theorem 3 establishes that only the conformal invariance   
is admitted. Its generators have the same form as in eq.   
\eqr{MFGCN}, where the summation is over the positive integers.

Finally, for the $D_{\infty}$ algebra one has
\[
\tilde{H}=\left(   
\begin{array}{ll}
-1 & -1 \\   
1 & -1
\end{array}
\right) =-\tilde{K}^{T}.
\]
Theorem 3 implies that only conformal transformations   
leave the system invariant and their generators are obtained
 by taking the limit
$N \rightarrow \infty$ in the formulas \eqr{confDN}.

\noindent{\bf 2. The semi-infinite Toda field theory \eqr{1.3}}
   
The discussion is very simple. Indeed, since $H$ is the indentity matrix,
 there   
are no gauge transformations. Moreover, the generators of   
the conformal transformations in the infinite, semi-infinite
and finite cases take always the same form \eqr{LezSavsymm}, where the   
summations
are over the appropriate range.

\noindent{\bf 3. The semi-infinite Toda field   
theories \eqr{1.4}}

 As opposed to the finite case, the matrix $H$ is no longer
 invertible, so now   
these theories are not equivalent to the   
ones given by \eqr{1.3}.   

First, we observe that, since $K$ is the identity matrix, there are no   
$\hat Z$-type transformations. For any classical Lie algebra,
 extended to $N \rightarrow \infty$, the recursive part of the
 systems, i.e. the equations labeled by $n \geq N_{0}$ as   
defined in Sec. 3.2, are always the same as in the infinite   
case discussed in Sec. 2.2.3. The solution of the   
corresponding difference equations  for $B_{n} \; \lf( n
  \geq N_{0} \ri)$ , that is \eqr{diff-1} and \eqr{diff-3},
are the same as in \eqr{solBilGer} and the generators   
are as in \eqr{symmBilGer}. However,   
for   
$ 1 \leq n < N_{0}$  the equations provide  constraints of   
the form   
\eqr{HtildeBtildea} and \eqr{HtildeBtildeb}. The   
application of the Theorem 3 implies   
\begin{itemize}
\item[1)] All the semi-infinite systems \eqr{1.4} are   
conformally invariant.
\item[2)] All the semi-infinite systems \eqr{1.4}   
have $n_{0} =1$, as defined in \eqr{ker}, hence a gauge
 transformation algebra of the form \eqr{semigauge} exists,
 with $j = 1$.   
\end{itemize}

In the $A_{\infty +}$ case the $\hat X$ and $\hat Y$   
conformal symmetries survive as in eq. \eqr{symmBilGer}, and so do   
${\hat U}_{2}$ and  ${\hat V}_{2}$ do. However the symmetries    
${\hat U}_{1}$ and  ${\hat V}_{1}$ are no longer present.

In the $B_{\infty}$ case the generators ${\hat X} , \;{\hat Y}$  and    
${\hat U}_{2}, \; {\hat V}_{2}$ combine together to give the new
conformal symmetry generators
\begin{equation}
\hat{X}=\xi \left( x\right) \partial _{x}+\frac{1}{2}\xi
_{x}\sum_{n=1}^{\infty }n\left( n-1\right) \partial _{u_{n}}, \qquad
\hat{Y}=\eta \left( y\right) \partial _{y}+\frac{1}{2}\eta
_{y}\sum_{n=1}^{\infty }n\left( n-1\right) \partial _{u_{n}}.
\end{equation}
The remaining gauge invariance is generated by   
\begin{equation}
\hat{U}\left( r\right) =r\left( x\right) \left[ \partial
_{u_{1}}+2\sum_{n=2}^{\infty }\partial _{u_{n}}\right], \qquad   
\hat{V}\left( s\right) =s\left( y\right) \left[ \partial
_{u_{1}}+2\sum_{n=2}^{\infty }\partial _{u_{n}}\right].   
\label{4.12}
\end{equation}

For the $C_{\infty}$ algebra the symmetry algebra is   
\begin{equation}
\begin{array}{l}
\ds{\hat{X}=\xi \left( x\right) \partial _{x}+\frac{1}{2}\xi_{x}
\sum_{n=1}^{\infty }n\left( n-2\right) \partial _{u_{n}},} \\   
\ds{\hat{Y}=\eta \left( y\right) \partial _{y}+\frac{1}{2}\eta_{y}
\sum_{n=1}^{\infty }n\left( n-2\right) \partial _{u_{n}},} \\   
\ds{\hat{U}\left( r\right) =r\left( x\right) \sum_{n=1}^{\infty }
\partial_{u_{n}},\quad \hat{V}\left( s\right) =s\left( y\right)
 \sum_{n=1}^{\infty
}\partial _{u_{n}}.}
\end{array}
\end{equation}
   
Finally, for the $D_{\infty}$ algebra one has
\begin{equation}
\begin{array}{l}
\ds{\hat{X}=\xi \left( x\right) \partial _{x}+\frac{1}{2}\xi_{x}
\sum_{n=1}^{\infty }\left( n-1\right) \left( n-2\right) 
\partial_{u_{n}},} \\ \\
\ds{\hat{Y}=\eta \left( y\right) \partial _{y}+\frac{1}{2}\eta_{y}
\sum_{n=1}^{\infty }\left( n-1\right) \left( n-2\right) \partial_{u_{n}},}
\\   \\
\ds{\hat{U}\left( r\right) =r\left( x\right) \left[
 \partial _{u_{1}}+\partial_{u_{2}}+2\sum_{n=3}^{\infty }\partial _{u_{n}}
 \right],}\\\\
\ds{\hat{V}\left(s\right) =s\left( y\right) \left[ \partial _{u_{1}}+
\partial_{u_{2}}+2\sum_{n=3}^{\infty }\partial _{u_{n}}\right].}
\end{array}
\label{4.14}
\end{equation}
The formulas for the semi-infinite  models \eqr{1.4}
are consistent with those obtained in the finite case in   
Sec. 3.2.3.   
The generators of the conformal invariance, in each case,
 are simply obtainable by dropping all terms involving $N$.
 Conversely, the terms proportional to a power of $N$ provide
 us with the gauge invariance generators in the semi-infinite   
 extensions. In this limit, the functions   
$r = \xi_{x}$ and
 $s = \eta_{y}$ must be considered as new linearly   
independent functions.    

\section{{\ Conclusions }}

We have introduced the generalized Toda system \eqr{1.1}
 and investigated its Lie point symmetry group. It turned
 out that in the infinite case $\lf( - \infty < n < \infty \ri)$
 these systems are always  invariant under an infinite   
dimensional group of conformal transformations. It is
 also gauge invariant, if a certain homogeneous linear   
difference equation (i.e. eq. \eqr{diff-1}) has
non trivial solutions. Further gauge transformations exist
 if another linear homogeneous difference equation
 (i.e. eq. \eqr{diff-2}) also has nontrivial solutions.

If we restrict the range of $n$ to $1 \leq n < \infty$,
 in some cases the symmetry group remains the same, or is reduced   
to a subgroup of the original symmetry group. However, 
in other cases (see   
\eqr{4.12} and \eqr{4.14}) the symmetry group does not coincide with a Lie 
subgroup.

In the finite case, with $1 \leq n \leq N$, the symmetry group
 remains the same as in the semi-infinite case, or it is    
reduced further.

In some situations (see Theorem 2 and 3) the infinite dimensional conformal
symmetry group is reduced to the Poincar\'e group in two   
dimensions (see eq. \eqr{Poincare}).

These results were obtained directly, that is  by analyzing the determining   
equations for the symmetries   
for all types of systems: infinite, semi-infinite and finite. The question 
to   
which we plan to devote a separate article is the
application of the infinite   
generalized Toda systems. In particular we will
establish the degree to which   
the symmetries of the semi-infinite and finite 
Toda systems are ``inherited''   
from those of the infinite systems. In other 
words we plan to discuss symmetry   
breaking by boundary or periodicity conditions 
of the infinite chains.   

One of the surprising results obtained in the present work   
is that the class of the conformally invariant Toda field
 theories is much larger than the class of the completely   
integrable models. Indeed, the existence of a Lax pair
imposes severe algebraic restrictions on the matrices   
$H$ and $K$ (see for instance \cite{19}).

\section*{\protect\bigskip \noindent {\sl Acknowledgments}      
\protect\medskip
\noindent}
This work is part of a project supported by the NATO CRG
960717, by the Italian INFN and by the
project SINTESI of the Italian Ministry  of the 
University and the Scientific   
Research. L.M. would like   
to thank
the Centre de Recherches Math\'ematiques of the Universit\'e de
 Montr\'eal for its warm
hospitality. S.L. and P.W. would like to thank the
Dipartimento di Fisica - Universit\'a di Lecce for its
 hospitality. The research of P.W. was partly supported by grants from NSERC   
and from FCAR. S.L.   
acknowledges a PhD scholarship from FCAR.

\pagebreak

\end{document}